\documentstyle[12pt,psfig]{article}
\def\bi{\bibitem}
\newcommand{\be}{\begin{equation}}
\newcommand{\ee}{\end{equation}}
\newcommand{\beq}{\begin{eqnarray}}
\newcommand{\eeq}{\end{eqnarray}}
\newcommand{\bear}{\begin{array}}
\newcommand{\ear}{\end{array}}

\newcommand{\bec}{\begin{center}}
\newcommand{\eec}{\end{center}}
\begin{document}
\title{Neutrino Oscillation as Coupled vibration.}    
\author{
Jyotisekhar Bhattacharyya\\ 
Dept. of Physics, Kanchrapara College,West Bengal,\\
P.O. Kanchrapara, India, \\
Satyabrata Biswas\\
Dept. of Physics, University of Kalyani\\
P. O. Kalyani, India, Pin.-741235  }
\date{}
\maketitle
\begin{abstract}
Neutrino oscillation is visualized as coupled vibrations. Unlike existing models describing neutrino oscillations, our model involving a single fundamental mass parameter $m$ and first discussed in the context of a possible boson oscillation demonstrates that for a self consistent theory there should be three types of bosons paired with maximal mixing. The argument is easily extended to neutrinos (fermions) with the exception that in the case of boson oscillation means periodic variation of particle number, whereas in the case of fermions the particle number does not change, only the mass oscillates. The fermionic vacuum is not really empty but filled with zero energy fermions. The oscillation length calculated perturbatively conforms to the experimental findings. Also there is no fermionic number violation in our model. 
\end{abstract} 
Though the neutrino oscillation has been a matter of interest for quite a long time, a satisfactory theory is yet to emerge because of the involvment of too many parameters \cite{bp:jetp,smb:pr,ca:prl,smb1:rmp,lb:prl,bz:ppn,moha:uni,gc:gu,gsl:prl,ksb:prl,rk:prd,rmp:astro}. In this letter we present a model involving a single fundamental mass parameter $m$, to explain the neutrino oscillation by identifying two different types of neutrinos having the same momentum with two simple harmonic oscillators stiffness coupled to one another.
\par
To get an insight into our model, we consider two real scalar fields $\phi_1$ and $\phi_2$ each of mass $m$ satisfying the Klein-Gordon equation
\be
(\Box+m^2)\phi_1=(\Box+m^2)\phi_2=0.
\ee
We take,
\beq
\phi_1&=&\int\frac{d^3k}{(2\pi)^32\omega_k}\left(a_1(k)e^{-ik.x}+a^\dagger_1(k)e^{ik.x}\right),\nonumber\\
\phi_2&=&\int\frac{d^3k}{(2\pi)^32\omega_k}\left(a_2(k)e^{-ik.x}+a^\dagger_2(k)e^{ik.x}\right),
\eeq
where $\omega_k=+\sqrt{k^2+m^2}$. Let us define a new scalar field $\phi=cos\theta\phi_1+sin\theta\phi_2$, satisfying $(\Box+m^2)\phi=0$. We take the mixing angle $\theta$ to be $45^0$, otherwise there will be inconsistency as explained later. Hence,
\be
\phi=\int\frac{d^3k}{(2\pi)^32\omega_k}\left(a(k)e^{-ik.x}+a^\dagger(k)e^{ik.x}\right).
\ee
The annihilation operators $a, a_1, a_2$ for $\phi, \phi_1, \phi_2$ respectively satisfy the following relation:
\be
a=\frac{a_1}{\sqrt{2}}+\frac{a_2}{\sqrt{2}}.
\ee
If we treat $\phi$ as a fundamental field, $\phi_1$ and $\phi_2$ lose their fundamental status (because the degree of freedom will be two instead of one). So we have to introduce three free fields $\chi_1\sim \phi_1+\phi_2,\,\,\chi_2\sim \phi_2+\phi_3,\,\,\chi_3\sim \phi_3+\phi_1$ to preserve the degrees of freedom and identify $\phi$ with any one of them, say $\chi_1$. We take $<\chi_i\vert \chi_j>=\delta_{ij}$ so that, $<\phi_i\vert \phi_j>\neq \delta_{ij}$. Hence there is possibility of oscillation. 
\par
In our qualitative discussion we now replace integration by summation over $k$ and put all numerical factors including $\omega_k $ as unity, only the factor $i=\sqrt{-1}$ is shown explicitly. This simplifies the Hamiltonian of the system after substracting the vacuum energy as
\be
H=\sum_k a^\dagger_ka_k\omega_k=\sum_k a^\dagger_ka_k,
\ee
where $a^\dagger_ka_k=N_k$ is the number operator for the state $k$ and the following commutation relations between $a$ and $a^\dagger$ are assumed:
\be
[a_k,a_{k^\prime}]=0,
\ee
\be 
[a_k,a_{k^\prime}^\dagger]=\delta_{k k^\prime}.
\ee
The above relations are satisfied if the following commutation relations hold:
\be
[a_{ik},a_{jk^\prime}^\dagger]=\delta_{ij}\delta_{kk^\prime},
\ee
\be
[a_{ik},a_{jk^\prime}]=0.
\ee
Thus the Hamiltonian of the system becomes
\be
H=\sum\left(a^\dagger_{1k}a_{1k}+a^\dagger_{2k}a_{2k}+a^\dagger_{1k}a_{2k}+a^\dagger_{2k}a_{1k}\right).
\ee
Now we define as usual the coordinates $Q_{ik}$ and the conjugate momenta $P_{ik}$ with $i=1,2$ as follows:
\beq
P_{ik}&=& (a_{ik}+a^\dagger_{ik})\nonumber\\
Q_{ik}&=& i(a_{ik}-a^\dagger_{ik}).
\eeq
Obviously $Q_{ik}, P_{ik}$'s are Hermitian operators and satisfies the canonical commutation relations 
\be
[Q_{ik},P_{jk^\prime}]= i\delta_{ij}\delta_{kk^\prime}, \,\,[Q_{ik},Q_{jk^\prime}]=[P_{ik},P_{jk^\prime}]=0, 
\ee
\par
If the mixing angle is $\theta$, $a_1,\,a_2$ should be replaced by $a_1cos\theta,\,a_2sin\theta$ respectively. 

 Hence,
\beq
Q_{1k}& = & cos\theta(a_{1k}+a_{1k}^\dagger),\nonumber\\
P_{1k}& = & icos\theta(a_{1k}-a_{1k}^\dagger)\nonumber\\
Q_{2k}& = & sin\theta(a_{2k}+a_{2k}^\dagger)\nonumber\\
P_{2k}& = & isin\theta(a_{2k}-a_{2k}^\dagger)
\eeq
So $[Q_{1k}, P_{1k}]=icos^2\theta,\,\,[Q_{2k},P_{2k}]=isin^2\theta$ but $[Q_{1k},P_{1k}]=[Q_{2k},P_{2k}].$ Hence $tan^2\theta=1$ or $\theta=45^0$. Also, the Hamiltonian for the state $k$ is $H_k=(a_{1k}^\dagger a_{1k}cos^2\theta+a_{2k}^\dagger a_{2k}sin^2\theta)\omega_k$. 
So the natural frequency of the systems (1) and (2) will be $cos^2\theta \omega_k$ and $sin^2\theta\omega_k$ respectively. If we replace $a_1$ by $a_1/{cos\theta}$ and $a_2/{cos\theta}$, $\phi$ will remain unchanged upto a normalization factor. The natural frequencies of the systems (1) and (2) become $\omega_k$ and $tan^2\theta\omega_k$. Thus to make them both equal to $\omega_k$, $tan^2\theta$ should be equal to unity i.e, $\theta=45^0$. This will be in accordance with the predictions of the existing theories describing neutrino oscillations that mixing is maximal for equal neutrino masses when we consider fermions later. From eqn.(10) 
\be
 H= \sum\left(Q_{1k}^2+P_{1k}^2+Q_{2k}^2+P_{2k}^2+4Q_{1k}Q_{2k}\right). 
\ee
\be
H_{ik}=Q_{ik}^2+P_{ik}^2=a_{ik}^\dagger a_{ik}\omega_k=N_{ik}\omega_k
\ee
 Equation (15) represents the free Hamiltonian of the $i-$th field for the state $k$ where $N_{ik}$ is the corresponding number operator.
Here, we have put 
\be
 a_{1k}a_{2k}=0.
\ee
When treated classically equation (14) represents stiffness coupled vibration of simple harmonic oscillators $1$ and $2$ with natural frequencies $\omega_{1k}=\omega_{2k}=2$ and coupling constant $\alpha=4$ for each $k$. The normal frequencies of the coupled system will be
\be
\omega_{+k}=\left[\frac{1}{2}(\omega_{1k}^2+\omega_{2k}^2)+\frac{1}{2}\sqrt{(\omega_{1k}^2-\omega_{2k}^2)^2+4\alpha^2}\right]^{1/2}=2\sqrt 2,
\ee
\be
\omega_{-k}=\left[\frac{1}{2}(\omega_{1k}^2+\omega_{2k}^2)-\frac{1}{2}\sqrt{(\omega_{1k}^2-\omega_{2k}^2)^2+4\alpha^2}\right]^{1/2}=0.
\ee
So, the actual frequency of oscillation will be $(\omega_{+k}+\omega_{-k})/2=\omega_{+k}/2=\sqrt{2}=\omega_k$,  after normalizing the fields $\phi_1$ and $\phi_2$ properly. The energy of each quantum oscillators characterized by $H_{ik}$, which is essentially  the corresponding particle number $N_{ik}$ (eqn. (15)) will oscillate with a frequency 
\be
\omega_{+k}=2\omega_k,
\ee
qualitatively, one type of particles changing into the other type alternately for each $k$.
\par
Eqn.(16) needs some explanation. Here $\phi$ is treated as the free scalar and  the Fock space is defined with respect to $\phi$, not $\phi_1,\phi_2$; $a_1^\dagger a_2$ destroys a particle of type 2 with the creation of a particle of type 1. This is exactly what happens in coupled oscillation. Therefore the term $a_1^\dagger a_2+ a_2^\dagger a_1$ in eqn(10) cannot be a priori put equal to zero. But the operator $a_1a_2$ destroys one particle of both types, which violates the law of conservation of energy and is dropped from the Hamiltonian (see eqn.(16)).    
\par
If we take complex field the number of degrees of freedom will only increase due to the addition of antiparticles.
\par Having discussed the toy model for boson, we switch to fermions. Let us consider two real fermion fields $\psi_1,\psi_2$ of the same mass $m$ satisfying the Dirac equation
\be
(i\gamma^\mu\partial_\mu-m)\psi_1=(i\gamma^\mu\partial_\mu-m)\psi_2=0.
\ee
Then $\psi=\frac{1}{\sqrt{2}}\psi_1+\frac{1}{\sqrt{2}}\psi_2$ will satisfy the same equation $(i\gamma^\mu\partial_\mu-m)\psi=0$. To preserve the number of degrees of freedom we three fields $\chi_1\sim \psi_1+\psi_2,\,\,\chi_2\sim \psi_2+\psi_3,\,\,\chi_3\sim \psi_3+\psi_1$ and identify $\psi$ with any one of them, say $\chi_1$ as in the scalar field case. The annihilation operators $a, a_1, a_2$ for $\psi, \psi_1, \psi_2$ respectively will satisfy the anti-commutation rules
\be
\{a_k,a_{k^\prime}\}=0,\,\,\,\,\{a_k,a_{k^\prime}^\dagger\}=\delta_{kk^\prime}
\ee
\be
\begin{array}{c}
\{a_{ik},a_{jk^\prime}\}=0,\nonumber\\
\{a_{ik},a_{jk^\prime}^\dagger\}+\{a_{jk},a_{ik^\prime}^\dagger\}=2\delta_{ij}\delta_{kk^\prime}+(1-2\delta_{ij})f_{kk^\prime},\,\,\,\, (i=1,2),
\end{array} 
\ee

where $f_{kk^\prime}$ is a suitable function of $k,k^\prime$. 

We define the coordinate $Q_{ik}$ and the conjugate momentum $P_{ik} $ in the same way as in Eqn. (11).
Inverting (11), we write
\beq
a_{ik}^\dagger&=& (P_{ik}+iQ_{ik})\nonumber\\
a_{ik}&=& (P_{ik}-iQ_{ik})
\eeq
and we impose the following subsidary conditions to satisfy (12):
\be
[a_{ik},a_{jk^\prime}]=0,\,\,\,\,[a_{ik},a^\dagger_{jk^\prime}]=\delta_{ij}\delta_{kk^\prime}
\ee
This gives
\be
\{a_{ik},a_{ik}^\dagger\}=Q_{ik}^2+P_{ik}^2=H_{ik}=a_{ik}^\dagger a_{ik}=1
\ee
 
This simply states that the number $N_{ik}$ of i-th fermion in  a particular state $k$ is constant (unity), i.e., the vacuum state is excluded from our consideration.  
The Hamiltonians will be given by the same equations (14), (15).
 It is essential that $f_{kk}\neq 0$ to make $Q_{1k}Q_{2k}\neq 0$. The system will be constrained otherwise unlike the scalar field case. So the energy of each system characterized by the Hamiltonian $H_{ik}= N_{ik}\omega_k$ will oscillate with a frequency $2\omega_k$, eqn. (19). Now, the particle number $N_{ik}$ is fixed ( =1 ), means $\omega_k$, hence the mass of each type of particle will oscillate with the same frequency, when one is maximum the other is minimum like sea-saw mechanism [13,14].   Actually for each $k$ the reaction force of the second oscillator on the first can be treated as a harmonic perturbation of frequency equal $\sim \omega_k$. This causes transition of the first oscillator between the successive energy levels ($0, \omega_k, 2\omega_k etc.)$ separated by the energy interval $\omega_k$ to the lowest order perturbation. Each state is a metastable state with life time $\Delta t\sim 1/\omega_k$ and so is not an eigenstate of energy (mass), the uncertainty in the energy being proportional to $1/{\Delta t}\sim \omega_k$. 
\par 
From the equation $a_{ik}^\dagger a_{ik}=1$ ( eqn.(25) ) all the fermion states are occupied, fermions are not really  annihilated, interactions can merely cause a fermion to jump to zero energy state and become a physical non-entity. the zero energy ground state of the harmonic oscillator, identified with the fermionic vacuum is therefore not really empty. This gives a new interpretation of fermionic vacuum.
 If we identify the first system with either of $\nu_e, \nu_\mu,\nu_\tau$, the second with another, the model gives a simple explanation of neutrino oscillation and the number of neutrino species being equal to three. Since all fermion states are occupied there is no fermion number violation. It also shows that $\nu_e, \nu_\mu,\nu_\tau$ are not mass eigenstates in accordance with the existing theories.
\par
We have treated the oscillators classically so far. To calculate the oscillation probability we turn to quantum mechanics and try to estimate oscillation length perturbatively. From eqn. (14) the first system is subjected to a perturbation Hamiltonian $H^\prime_{1k}\sim \omega_k^2Q_{1k}Q_{2k}$ (keeping track of the factors $\omega_k $), where $Q_{2k}\sim (1/{\sqrt{\omega_k}})e^{i2\omega_k t}$ according to classical equations of motion. To lowest order, this harmonic perturbation causes transition of the first system from the first excited to the ground state with probability
\beq
&\sim& \omega_{k}^3\left(\int_{-\infty}^{+\infty}u_f(Q_{1k})Q_{1k}u_i(Q_{1k})dQ_{1k}\right)^2sin^2(\omega_{fi}+2\omega_k)t/{(\omega_{fi}+2\omega_{k})^2}\nonumber\\
&\sim & \omega_{k}^2sin^2(\omega_{fi}+2\omega_{k})t/{(\omega_{fi}+2\omega_{k})^2}
\eeq 
according to time dependent perturbation theory where $u_f=u_0(Q_{1k}),u_i=u_1(Q_{1k})$ are the wave functions for the ground and the first excited state of a harmonic oscillator of frequency $\omega_k$ respectively. So $\omega_f=0, \omega_i=\omega_k$ and $\omega_{fi}+2\omega_k=\omega_{k}$. Thus the transition probability becomes proportinal to $sin^2(\omega_k t)$. Similarly the probability of transition of the second system from the ground to the first excited state with energy $\omega_{k^\prime}$ is proportional to $sin^2\omega_{k^\prime}t$. If $k\neq k^\prime$ the first and the second systems will belong to two independent coupled oscillators and the two transitions will be independent of one another. Thus the probability, that a neutrino of type 1 of energy $\omega_k$ is annihilated and at the same time a neutrino of energy $\omega_{k^\prime}$ of type 2 is created, will be the product of the above two probabilities and will be given by
\be
P_{kk^\prime}\sim sin^2\omega_k tsin^2\omega_{k^\prime}t.
\ee
When we consider aneutrino beam $t$ can be replaced by the distance $x$ travelled by the beam, since neutrinos are extremely relativistic. Thus we get 
\beq
P_{kk^\prime}& \sim& [cos(\omega_{k^\prime}-\omega_k)x-cos(\omega_{k^\prime}+\omega_{k})x]^2\nonumber\\
& = & cos^2(\omega_{k^\prime}+\omega_k)x-2cos(\omega_{k^\prime}+\omega_k)xcos(\omega_{k^\prime}-\omega_k)x+cos^2(\omega_{k^\prime}-\omega_k)x\nonumber\\
\eeq 
From eqn. (28), if $\omega_{k^\prime}>>\omega_k, P_{kk^\prime}\simeq 0$. So if neutrinos differ widely in energy the oscillation probability becomes extremely small. On the otherhand if $\omega_{k^\prime}\sim \omega_k\simeq 1$ GeV as for $\nu_\mu\rightarrow \nu_\tau$ oscillation the oscillation length corresponding to the frequency $\omega_k$ will be much less than the uncertainty in the position of the detector. So $cos^2(\omega_{k^\prime}+\omega_k)t$ and $cos(\omega_{k^\prime}+\omega_k)t$ in eqn. (28) should be replaced by $1/2$ and $0$ respectively for distances much higher than $1GeV^{-1}$. Thus 
\be
P_{kk^\prime}\sim \frac{1}{2}+cos^2(\omega_{k^\prime}-\omega_k)x
\ee
which corresponds to oscillation length 
\beq
L\sim 1/{(\omega_{k^\prime}-\omega_k)}&=&\frac{\omega_{k^\prime}+\omega_k}{\omega_{k^\prime}^2-\omega_{k}^2}\nonumber\\
&\sim & \frac{\omega_k}{\omega_{k^\prime}^2-\omega_{k}^2}
\eeq
For $\omega_{k^\prime}^2-\omega_k^2\sim 10^{-3}$eV this gives $L\sim 1000$km, in qualitative agreement with the experimental results \cite{ys:iv}. 
  
\par
To conclude, from the expression for the free particle Hamiltonian (eqn.(20) it is evident that particle states are represented by excitations of simple harmonically oscillating systems. The true nature of the system can be revealed only when the system interacts with other systems. From the eqn.(14) the coupling between two such systems is like two springs coupled to one another. Thus invidual particle states should be represented by spring excitations.\\
{\bf Acknowledgement:}\\ The authors are thankful to G. Bhattacharyya and P. B. Pal for help during the preparation of the manuscript.\\ 
         
\end{document}